\documentclass[nofootinbib,aps,preprintnum,11pt,bers,unsortedaddress]{revtex4}
\usepackage{graphicx}
\usepackage{cancel}
\usepackage{amssymb}
\usepackage{textcomp}
\usepackage{feynmp-auto}
\usepackage{amsmath}
\usepackage{bm}
\usepackage{times}
\usepackage{epsfig}
\usepackage{color}
\usepackage{graphics}
\usepackage{hyperref}
\usepackage{setspace}
\usepackage{paralist}
\usepackage{amsmath}

\usepackage{tikz}

\usepackage{array,booktabs}

\usepackage{bbm}
\usepackage{soul}
\usepackage{xcolor}

\usepackage{epsfig}
\usepackage{color}
\usepackage{slashed}
\usepackage{comment}
\usepackage{epstopdf}
\usepackage{array}
\usepackage{tikz}

\usepackage{rotating}
\usepackage{framed}

\usepackage{placeins}

\usepackage{setspace}
\usepackage{paralist}
\usepackage{multirow}
\usepackage{amsmath}
\usepackage[usenames,dvipsnames]{pstricks}

\usepackage{amsmath} 
\usepackage{amsthm} 
\usepackage{amssymb}	
 
\newcommand{\matrixel}[3]{\left< #1 \vphantom{#2#3} \right|
 #2 \left| #3 \vphantom{#1#2} \right>} 
\let\baraccent=\= 
\renewcommand{\=}[1]{\stackrel{#1}{=}} 

\theoremstyle{definition}

\theoremstyle{remark}

\usepackage{bbm}

\hypersetup{
    pdfnewwindow=true,      
    colorlinks=true,       
    linkcolor=black,          
    citecolor=blue,        
    filecolor=blue,      
    urlcolor=blue           
}

\makeatletter
\newcommand\xleftrightarrow[2][]{%
  \ext@arrow 9999{\longleftrightarrowfill@}{#1}{#2}}
\newcommand\longleftrightarrowfill@{%
  \arrowfill@\leftarrow\relbar\rightarrow}
\makeatother

\begin{document}
\title{\Large{Seesaw Scale, Unification, and Proton Decay}}
\author{Pavel Fileviez P\'erez$^{1}$, Axel Gross$^{1}$, Clara Murgui$^{2}$}
\affiliation{$^{1}$Physics Department and Center for Education and Research in Cosmology and Astrophysics (CERCA), Case Western Reserve University, Rockefeller Bldg. 2076 Adelbert Rd. Cleveland, OH 44106, USA \\
$^{2}$Departamento de F\'isica Te\'orica, IFIC, Universitat de Valencia-CSIC, 
E-46071, Valencia, Spain}
\email{pxf112@case.edu, apg37@case.edu, clara.murgui@ific.uv.es.}

\begin{abstract}
We investigate a simple realistic grand unified theory based on the $SU(5)$ gauge symmetry which predicts an upper bound on the proton decay lifetime for the channels $p \to K^+ \bar{\nu}$ and $p \to \pi^+ \bar{\nu}$, i.e. $\tau (p \to K^+ \bar{\nu}) \lesssim 3.4 \times 10^{35}$ and $\tau (p \to \pi^+ \bar{\nu}) \lesssim 1.7 \times 10^{34}$ years, respectively. In this context, the neutrino masses are generated through the type I and type III seesaw mechanisms, and one predicts that the field responsible for type III seesaw must be light with a mass below 500 TeV. We discuss the testability of this theory at current and future proton decay experiments.
\end{abstract}

\maketitle

\section{Introduction}
One of the main goals of theoretical physics is to understand the unification of fundamental forces in nature. In 1974, H. Georgi and S. Glashow proposed the simplest grand unified theory (GUT) ~\cite{Georgi:1974sy}, based on the $SU(5)$ gauge group, and it has since been considered as one of the most appealing extensions of the Standard Model of Particle Physics. One of the most impressive predictions of $SU(5)$ is the decay of the proton; for a review on proton decay, see Ref.~\cite{Nath:2006ut}. This theory could describe physics at the high scale, $M_{GUT} \sim 10^{14-15}$ GeV, but unfortunately, it is ruled out by experiment. The main problems of the Georgi-Glashow model are the following:
\begin{itemize}
\item {\textit{Gauge Coupling Unification}}: If one assumes the existence of the SU(5) theory at the high scale and studies the running of the gauge couplings, it is simple to show that the values of the SM gauge couplings cannot be reproduced at the electroweak scale. To rectify this, one can include new degrees of freedom which help to achieve unification in agreement with experiment.
\item {\textit{Charged Fermion Masses}}: The theory predicts that the masses of the charged leptons and down quarks are equal at the high scale, i.e. $Y_e=Y_d^T$. Unfortunately, this prediction cannot reproduce the observed values for their masses at the low scale. Conventionally, there are three ways to achieve a consistent relation between these masses: a) Include higher-dimensional operators suppressed by the Planck scale~\cite{Ellis:1979fg}, b) Add a new Higgs in the ${\bf{45}}$ representation~\cite{Georgi:1979df}, or c) Add new vector-like fermions.
The last possibility is the simplest, and in this case, the theory can be more predictive. We will discuss this possibility in detail.
\item {\textit{Neutrino Masses}}: As in the Standard Model, the neutrinos are massless in $SU(5)$. One can generate neutrino masses through the different seesaw mechanisms:
\begin{itemize}
\item Type I ~\cite{TypeI}: The masses can be generated by adding at least two copies of right-handed neutrinos, ${\bf{1}_i}$.
\item Type II~\cite{TypeII}: The masses can be generated by the addition of a new Higgs representation, ${\bf{15_H}}$. In this case, the field important for the seesaw mechanism, $\Delta\sim (1,3,1)$, lives in the ${\bf{15_H}}$ representation. See Ref.~\cite{Dorsner:2005fq} for a simple model that implements the type II seesaw mechanism.
\item Type III~\cite{TypeIII} and I: The masses can be generated by adding the fermionic ${\bf{24}}$ representation. In this case, the fields needed for type I and type III seesaw mechanisms are $\rho_0 \sim (1,1,0)$ and $\rho_3 \sim (1,3,0)$, respectively. See Refs.~\cite{Bajc:2006ia,Bajc:2007zf,Perez:2007rm,Dorsner:2006fx,Ma:1998dn} for the implementation of this mechanism.

\item Zee mechanism~\cite{Zee:1980ai}: One can generate neutrino masses at the one-loop level by adding two new Higgs fields: a charged singlet in ${\bf{10_H}}$ and a second Higgs doublet in ${\bf{45_H}}$.
See Ref.~\cite{Perez:2016qbo} for the implementation of this mechanism in a simple renormalizable model.

\end{itemize}

\end{itemize}

Following the above discussion, one can think about different realistic extensions of the Georgi-Glashow model. In this article, we investigate a simple renormalizable extension of the Georgi-Glashow model that corrects the three major problems with the Georgi-Glashow model: neutrino masses, consistent charged fermion masses, and unification of gauge couplings. In this theory, one can achieve a consistent relation between the charged fermion masses by adding vector-like fermions in the ${\bf{5^{'}}}$ and ${\bf{\bar{5}^{'}}}$ representations. The neutrino masses are generated through the type I and type III seesaw mechanisms, but in this case, the new vector-like fermions also play a crucial role. In this context, we show that we can achieve the unification of the gauge couplings in agreement with the low energy constraints. We find that the field generating neutrino masses through the type III seesaw mechanism must be light, i.e. $M_{\rho_3} < 500$ TeV. We discuss the predictions for proton decay and show that the theory predicts an upper bound on the proton decay lifetime for the channels $p \to K^+ \bar{\nu}$, i.e. $\tau (p \to K^+ \bar{\nu}) \lesssim  3.4 \times 10^{35}$ years, and $p \to \pi^+ \bar{\nu}$, i.e. $\tau(p \to \pi^+ \bar{\nu}) \lesssim  1.7 \times 10^{34}$ years. We discuss the constraints on the spectrum of the theory and the generation of neutrino masses in detail. The model proposed in this article can be considered as one of the most appealing realistic extensions of the Georgi-Glashow model.

This article is organized as follows: In section II, we discuss the main features of our model: the unification constraints, the generation of neutrino and charged fermion masses, and the predictions for proton decay. We summarize our main results in section III. 

\section{Theoretical Framework}
We focus on a simple renormalizable theory with the following properties:
\begin{itemize}
\item {\textit{Fermions}}: As in the original Georgi-Glashow model, we have the Standard Model fermions in the ${\bf{\bar{5}}}$ and ${\bf{10}}$ representations. We add
the ${\bf{24}}$ representation to generate neutrino mass and, in the spirit of the Standard Model, three copies of vector-like fermions in
${\bf{\bar{5}^{'}}}$ and ${\bf{5^{'}}}$ representations to achieve a realistic relation between the charged fermion masses. We list the representations to set our notation:
\begin{equation}
\bar{5}=\begin{pmatrix} d^c \\ \ell \end{pmatrix}, \quad 10=\begin{pmatrix} u^c && Q \\ Q
 && e^c \end{pmatrix}, \quad \bar{5}^{'}=\begin{pmatrix} D^c \\ L \end{pmatrix}, \quad 5^{'}=\begin{pmatrix} D \\ L^c \end{pmatrix}, \quad 24= \begin{pmatrix} \rho_8 && \rho_{(3,2)} \\ \rho_{(\bar{3},2)} && \rho_3 \end{pmatrix} + \lambda_{24} \rho_0, \nonumber
\end{equation}
where $\lambda_{24}=1/(2\sqrt{15})\,\text{Diag}(2,2,2,-3,-3)$.

\item {\textit{Gauge Bosons}}: The Standard Model gauge bosons live in the adjoint representation, and we use the following notation:
\begin{equation}
 A_{\mu}= \begin{pmatrix} G_\mu && V_\mu \\ V_\mu^\dagger && W_\mu \end{pmatrix} +  \lambda_{24} B_\mu. \nonumber
\end{equation}
\item {\textit{Scalar Sector}}: We stick to the minimal Higgs sector of the Georgi-Glashow model:
\begin{equation}
5_H=\begin{pmatrix} T \\ H \end{pmatrix}, \quad 24_H= \begin{pmatrix} \Sigma_8 && \Sigma_{(3,2)} \\ \Sigma_{(\bar{3},2)} && \Sigma_3 \end{pmatrix} + \lambda_{24} \Sigma_{24}. \nonumber
\end{equation}
\end{itemize}
Now we describe the splitting between the fields in the new fermionic representations:

\begin{itemize}

\item {\textit{New Seesaw Fields}}: We write the following terms relevant for the mass of the ${\bf{24}}$:
\begin{equation}
- {\cal L}_{24}\supset M \ \text{Tr}\{ 24^2\} + \lambda \ \text{Tr} \{ 24^2 24_H \} + \rm{h.c.}.
\end{equation}

After the grand unified symmetry is broken, the masses for the fields in the ${\bf{24}}$ representation are given by: 
\begin{align*}
&M_{\rho_8}= M + \tilde{\lambda} \, \frac{M_{GUT}}{\sqrt{\alpha_{GUT}}}, &M_{\rho_3}= M - \frac{3}{2} \, \tilde{\lambda} \, \frac{M_{GUT}}{\sqrt{\alpha_{GUT}}},\\
&M_{\rho_{(3,2)}}=M -\frac{1}{4} \,  \tilde{\lambda} \, \frac{M_{GUT}}{\sqrt{\alpha_{GUT}}},  &M_{\rho_0}=M-\frac{1}{2} \, \tilde{\lambda}\frac{M_{GUT}}{\sqrt{\alpha_{GUT}}},
\end{align*}
where $\tilde{\lambda}=\lambda/\sqrt{25\pi}$ and $M_{GUT}=\sqrt{5\pi \alpha_{GUT}/3} \, v_{24}$. Defining $\hat{m}_{24}=M_{\rho_8}/M_{\rho_3}$, the masses in the ${\bf{24}}$ can be defined as a function of $M_{\rho_3}$: 
\begin{equation}
M_{\rho_8}=\hat{m}_{24} M_{\rho_3}, \quad M_{\rho_0}=\frac{1}{5}(3+2 \, \hat{m}_{24})M_{\rho_3}, \quad M_{\rho_{(3,2)}}=M_{\rho_{(\bar{3},2)}}=\frac{1}{2}(1+\hat{m}_{24})M_{\rho_3}.
\end{equation}
\item $\bf{{\it 5}^{'}}$ {\textit{and}} ${\bf{\bar{{\it 5}}^{'}}}$ {\textit{fields}}: Using the following terms in the Lagrangian:
\begin{equation}
- {\cal L}_{5}\supset M_5 \, \bar{5}^{'}5^{'} + \lambda_5 \, \bar{5}^{'} 24_H 5^{'} + \rm{h.c.},
\end{equation}
we can find the masses of the fields in the ${\bf{5^{'}}}$ and ${\bf{\bar{5}^{'}}}$ 
fermionic representations:
\begin{align*}
&M_{D}= M_{5} + {\tilde{\lambda}}_5 \frac{M_{GUT}}{\sqrt{\alpha_{GUT}}}, &M_{L}^T= M_{5} - \frac{3}{2}{\tilde{\lambda}}_5\frac{M_{GUT}}{\sqrt{\alpha_{GUT}}},
\end{align*}
where ${\tilde{\lambda}}_5=\lambda_5 / \sqrt{25\pi}$.
In the same spirit as the splitting of the ${\bf 24}$, we can define $\hat{m}_{5} =M_{D}/M_{L}$ to write the mass of the down-type quarks as a function of the mass of the leptons.
\item {\textit{Yukawa couplings}}: The Yukawa terms relevant to understanding the generation of fermion masses are given by:
\begin{equation}
 \begin{split}
-{\cal L}_f \supset & \  {y_0^i} \,  \bar{5}_i  \, 24  \,5_H + {y_1^i} \, \bar{5}_i^{'} \, 24 \, 5_H + {y_2^i} \, 5_H^* \, 24 \, 5_i^{'} + Y_1 \, 5_H^* \,\bar{5} \, 10+Y_2	 \, 5_H^* \,\bar{5}^{'} \, 10 + Y_u \, 10 \, 10 \, 5_H \, \\
& + M_{\bar{5}5} \, \bar{5} \, 5^{'} + \lambda_{\bar{5}5} \, \bar{5} \, 24_H \,  5^{'} + \rm{h.c.}
\end{split}
\end{equation}
\end{itemize}
\subsection{Gauge Unification Constraints}
The pragmatic way to determine if one can achieve gauge coupling unification in agreement with the low energy constraints is to assume unification at the high scale and to constrain the full spectrum of the theory using the allowed freedom. The equations for the running of the gauge couplings are given by:
\begin{equation}
 \displaystyle \alpha_i^{-1}(M_Z) = \alpha^{-1}_{GUT}+\frac{B_i}{2 \pi} \ln \left( \frac{M_{GUT}}{M_Z} \right),
 \label{RGE_coupling}
\end{equation}
where
\begin{equation}
B_i=b_i^{SM} + b_{iI}\, r_I, \, \quad \ r_I= \frac{ \ln (M_{GUT} / M_I)}{\ln (M_{GUT} / M_Z)},
\end{equation}
and $M_I$ is the mass of any new particle living in the great desert. These equations can be rewritten in a more suitable form in terms of the differences of the coefficients and the low energy observables. Assuming unification, these equations can be reduced to:
\begin{eqnarray}
 \displaystyle \frac{B_{23}}{ B_{12}}&=&\frac{5}{8}\left( \frac{\text{sin}^2\theta_W (M_Z) - \alpha (M_Z) / \alpha_s (M_Z)}{ 3/8-\text{sin}^2\theta_W (M_Z)}\right), \\
 \label{ratio_unification} \nonumber \\
 \displaystyle \ln \left(\frac{M_{GUT}}{M_Z}\right)&=&\frac{16\pi }{ 5\alpha (M_Z)}\left(\frac{3/8-\sin^2\theta_W (M_Z) }{ B_{12}}\right), \label{gut_scale}
\end{eqnarray}
where $B_{ij} = B_i - B_j$. Using the experimental values $\alpha_s(M_Z)=0.1182$, $\alpha^{-1}(M_Z)=127.95$, and $\sin^2 \theta_W(M_Z) = 0.2313$~\cite{Patrignani:2016xqp}, we find:
\begin{equation}
 \frac{B_{23}}{B_{12}}= 0.718, \quad   \ln \left(\frac{M_{GUT}}{M_Z}\right)=\frac{184.84 }{ B_{12}}.
\end{equation}
These equations can be used to constrain the spectrum of the theory.

\begin{table}[h] 
  \vspace{0.5cm}
 \begin{tabular}{c | c  c | c c  | c  c | c  c c c c  }
  \hline \hline
  \multicolumn{3}{c}{\quad ${\bf{5_H}}$} &  \multicolumn{2}{c}{${\bf{24_H}}$} &  \multicolumn{2}{c}{${\bf{\bar{5}^{'}}}+{\bf{5^{'}}}$} &  \multicolumn{3}{c}{${\bf{24}}$} \\
  \hline \hline
   & $H$&  $T$ &  $\Sigma_3$  & $\Sigma_8$ & $L$ + $L^c$ & $D$ + $D^c$ & $\rho_3$ & $\rho_8$  & $\rho_{(3,2)} + \rho_{(\bar{3},2)}$ \\
\hline
& & & & & & & \\[-0.35cm]
$B_1$ & $\frac{1}{10}$ 	& $\frac{1}{15} \, r_T$	& $0$					& $0$					& $\frac{6}{5} \, r_L$		& $\frac{4}{5} \, r_D$	& $0$ 					& $0$			&  $\frac{10}{3} \, r_{32}$ \\[0.1cm]
$B_2$ & $\frac{1}{6} $ 	& $0$		& $\frac{1}{3} \, r_{\Sigma_3}$	& $0$					& $ 2 \, r_L$		& $0$				& $\frac{4}{3} \, r_{3}$		& $0$			& $2 \, r_{32}$ \\[0.1cm]
$B_3$ & $0$			& $\frac{1}{6}$	& $0$					& $\frac{1}{2}\, r_{\Sigma_8}$ 	& $0$				& $2 \, r_D$		& $0$					& $2 \, r_{8}$	& $\frac{4}{3} \, r_{32}$ \\[0.1cm]
\hline
& & & & & & & \\[-0.35cm]
$B_{12}$ & $-\frac{1}{15}$& $\frac{1}{15} \, r_T$& $-\frac{1}{3} \, r_{\Sigma_3}$	& $0$					& $-\frac{4}{5}\,  r_L$	& $\frac{4}{5}  \,r_D$	& $-\frac{4}{3} \, r_{3}$		& $0$ 			& $\frac{4}{3} \, r_{32}$ \\[0.1cm]
$B_{23}$ & $\frac{1}{6}$	& $-\frac{1}{6} \, r_T $ & $\frac{1}{3} \, r_{\Sigma_3}$	& $-\frac{1}{2} \, r_{\Sigma_8}$	&$2 \, r_L$		& $- 2 \, r_D$		& $\frac{4}{3} \, r_{3}$		& $-2 \, r_{8}$ 	& $\frac{2}{3} \, r_{32}$\\[0.1cm]
  \hline \hline
  \end{tabular}
 \label{B_coef_SU5}
 \caption{Contributions to the beta functions from the new particles in the theory.}
 \end{table}
In Table I, we list the contributions to the $B_{ij}$ coefficients in the theory. The relevant equations for our analysis can be explicitly written as:
\begin{eqnarray}
B_{12}&=&B_{12}^\text{SM}-\frac{4}{5} \, r_5-\frac{4}{3} \, r_{3}-\frac{1}{3} \, r_{\Sigma_3}+\frac{4}{3} \, r_{32},\\
B_{23}&=&B_{23}^\text{SM}+ 2 \,  r_5+\frac{4}{3} \, r_{3}+\frac{1}{3} \, r_{\Sigma_3}-2 \, r_{8}-\frac{1}{2} \, r_{\Sigma_8}+\frac{2}{3} \, r_{32},
\end{eqnarray}
where $B_{12}^{SM}=109/15$, $B_{23}^{SM}=23/6$ and $r_5 = r_L - r_D$. We assume that the colored triplet in the ${\bf 5_H}$ lives at the high scale because it mediates proton decay. We note that $r_5$ is only a function of the mass splitting $\hat{m}_5$, i.e. $\displaystyle r_5 = \ln \hat{m}_5 / (\ln M_{GUT} - \ln M_Z)$. Since unification is only sensitive to the splitting in the mass of the representations, we can eliminate the overall mass scales and write the above equations in a simple way:
\begin{eqnarray}
B_{12}&=&B_{12}^\text{SM}-\frac{4}{5} \, r_5-\frac{1}{3} \, r_{\Sigma_3}-\frac{4}{3} \, \frac{\ln\left(\frac{1}{2}(1+\hat{m}_{24})\right)}{\ln(M_{\text{GUT}})-\ln(M_Z)},\\
B_{23}&=&B_{23}^\text{SM}+ 2 \,  r_5+\frac{1}{3} \, r_{\Sigma_3}+2 \, \frac{\ln(\hat{m}_{24})}{\ln (M_\text{GUT})-\ln(M_Z)}-\frac{1}{2} \, r_{\Sigma_8}-\frac{2}{3} \, \frac{\ln\left(\frac{1}{2}(1+\hat{m}_{24})\right)}{\ln(M_{\text{GUT}})-\ln(M_Z)}.
\end{eqnarray}

\begin{figure}[ht]
\includegraphics[width=0.49\linewidth]{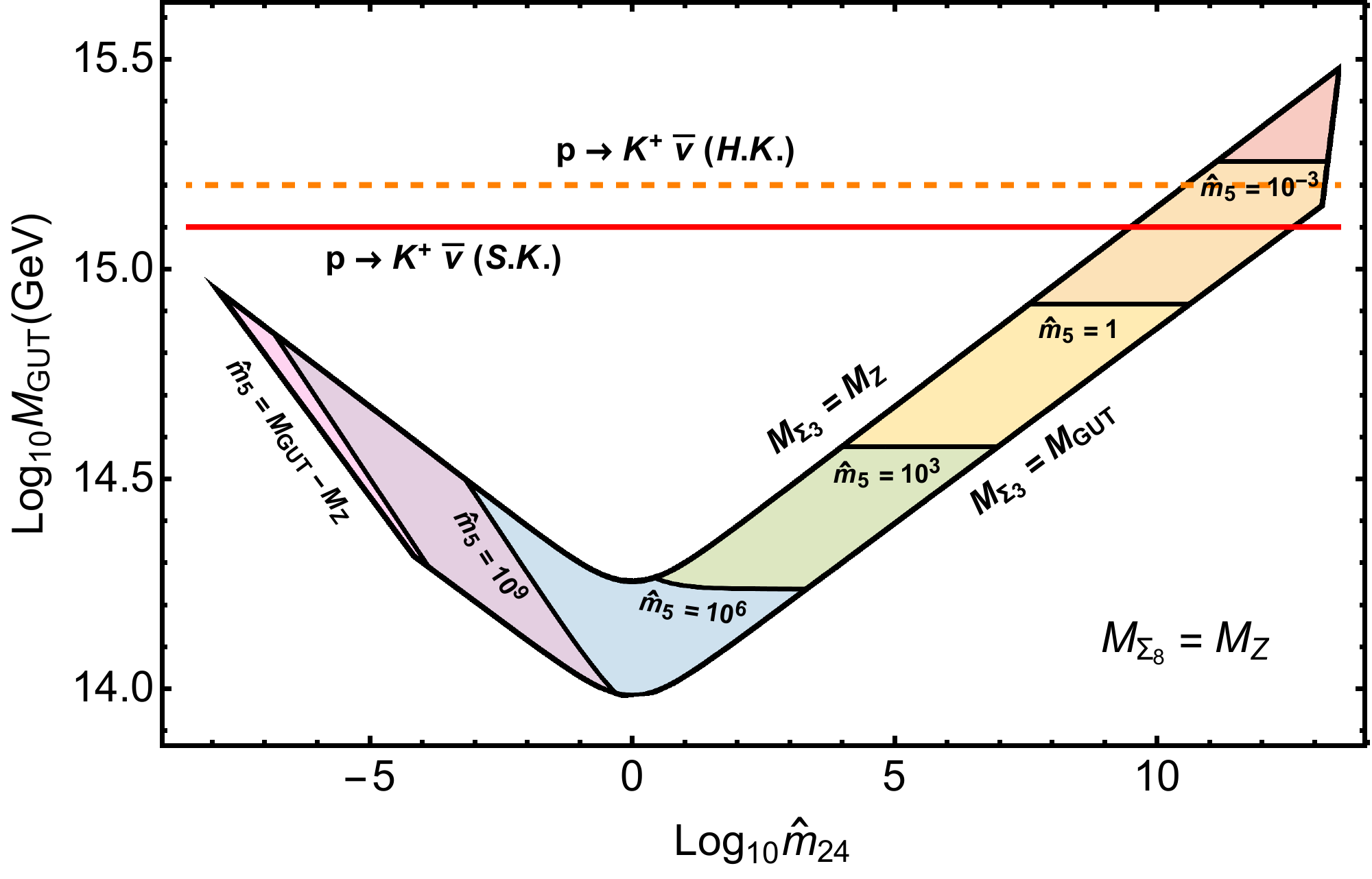}
\includegraphics[width=0.49\linewidth]{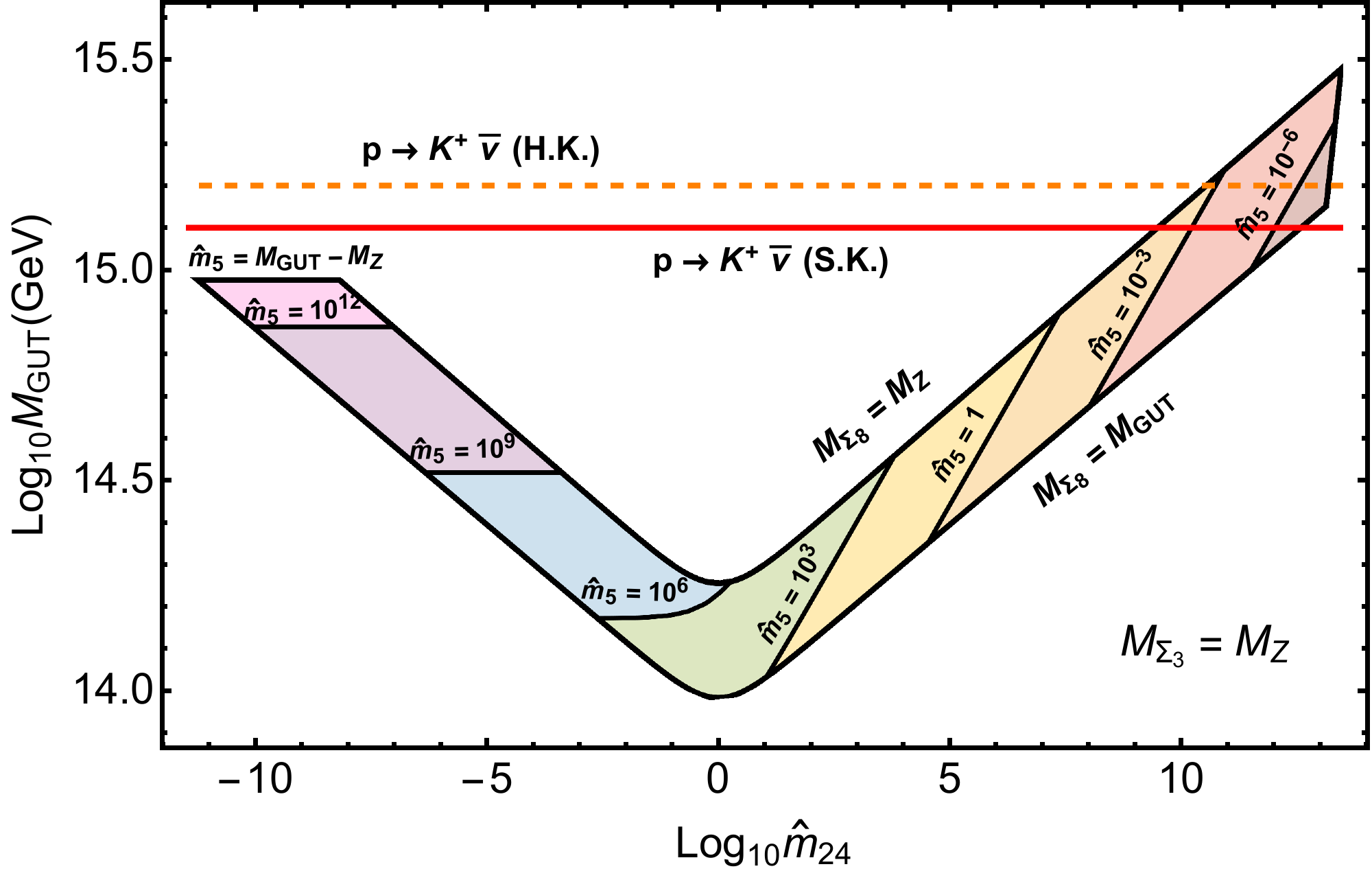}
\caption{Parameter space allowed by the unification of gauge couplings in the plane $\rm{Log}_{10} M_{GUT}-\rm{Log}_{10} \hat{m}_{24}$. Every point in the colored area corresponds to a scenario in which the gauge couplings unify at the high scale. In the left panel, we show the constraints on the mass spectrum when $M_{\Sigma_8}=M_{Z}$ while changing the mass of $\Sigma_3$ between the electroweak and the GUT scales. In the right panel, we show the same constraints in the case when $M_{\Sigma_3}=M_{Z}$ while changing the mass of $\Sigma_8$.The red horizontal line corresponds to the current bound on the proton decay lifetime for the channel $p \to K^+ \bar{\nu}$ \cite{Abe:2014mwa}, and the orange dashed line corresponds to the projected bound from the Hyper-Kamiokande collaboration \cite{Yokoyama:2017mnt}.}
\end{figure}
In Fig.1, we show the parameter space allowed by the unification of gauge couplings in the plane $\rm{Log}_{10} M_{GUT}-\rm{Log}_{10} \hat{m}_{24}$. We note that even though there is a large overall parameter space, the parameter space that gives unification compatible with proton decay is limited. Specifically, we find that $\hat{m}_{24}\gtrsim 10^{9}$, $\hat{m}_5 \lesssim 10^{-2}$, and that $\Sigma_8$ and $\Sigma_3$ should live near the electroweak scale. We find the maximum possible GUT scale is $10^{15.5}$ GeV. The bounds coming from proton decay experiments will be discussed in the next section. 

We have shown the unification and proton decay constraints on the mass splitting for the fermionic fields living in the ${\bf{24}}$, ${\bf{5}^{'}}$, and ${\bf{\bar{5}^{'}}}$ representations, but
it is also useful to explicitly show the allowed masses for these fields. In Fig.~2, we show the allowed parameter space for the masses of the new fermions for the most optimistic case: $M_{\Sigma_3}=M_{\Sigma_8}=M_Z$. We find that the seesaw field generating neutrino masses through the type III seesaw, $\rho_3$, has a mass at the multi-TeV scale, with an upper bound of $M_{\rho_3} \leq 500$ TeV. This is an interesting result which allows for the possibility to test the type III seesaw mechanism at current or future colliders.  

\begin{figure}[h]
\includegraphics[width=0.49\linewidth]{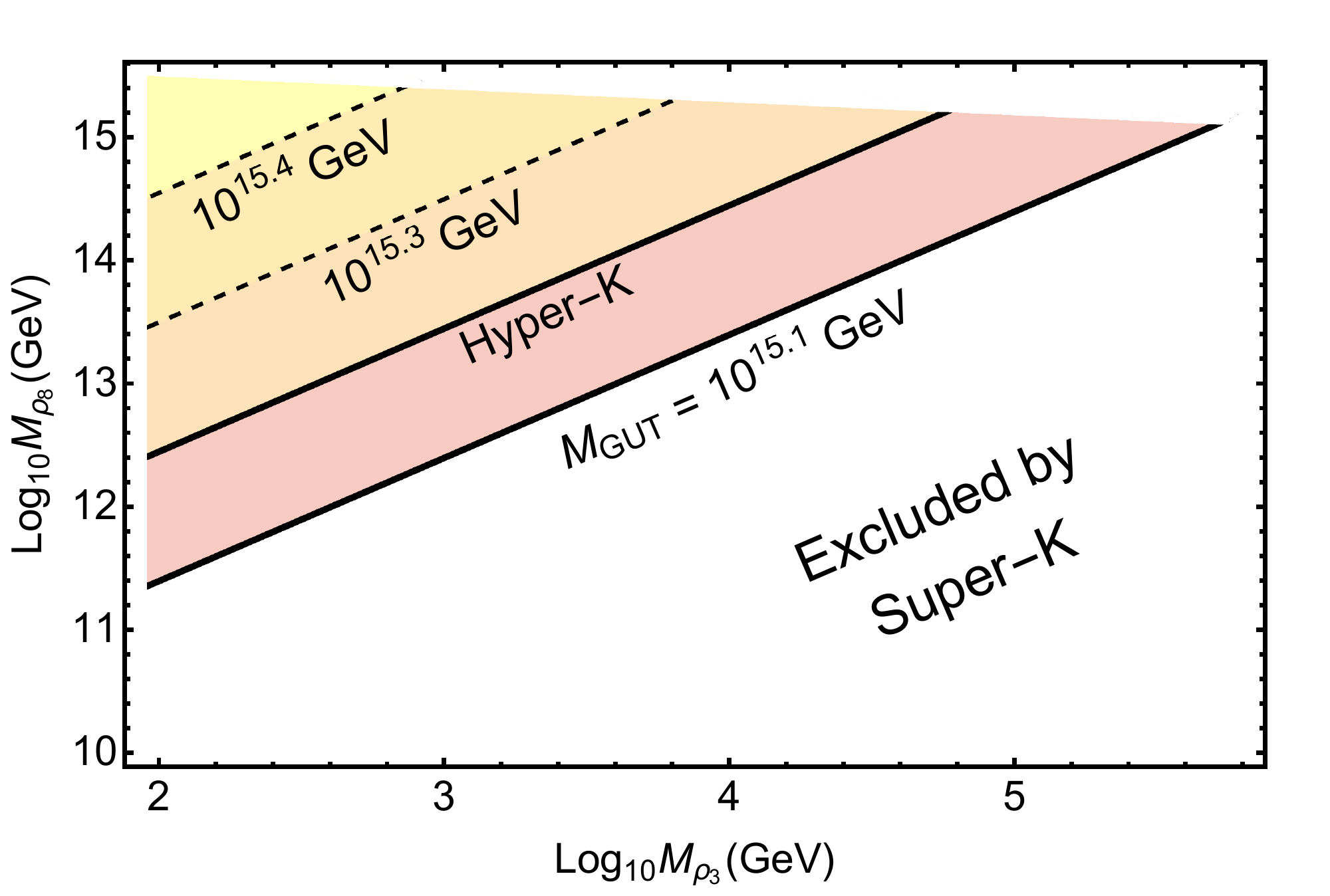}
\includegraphics[width=0.49\linewidth]{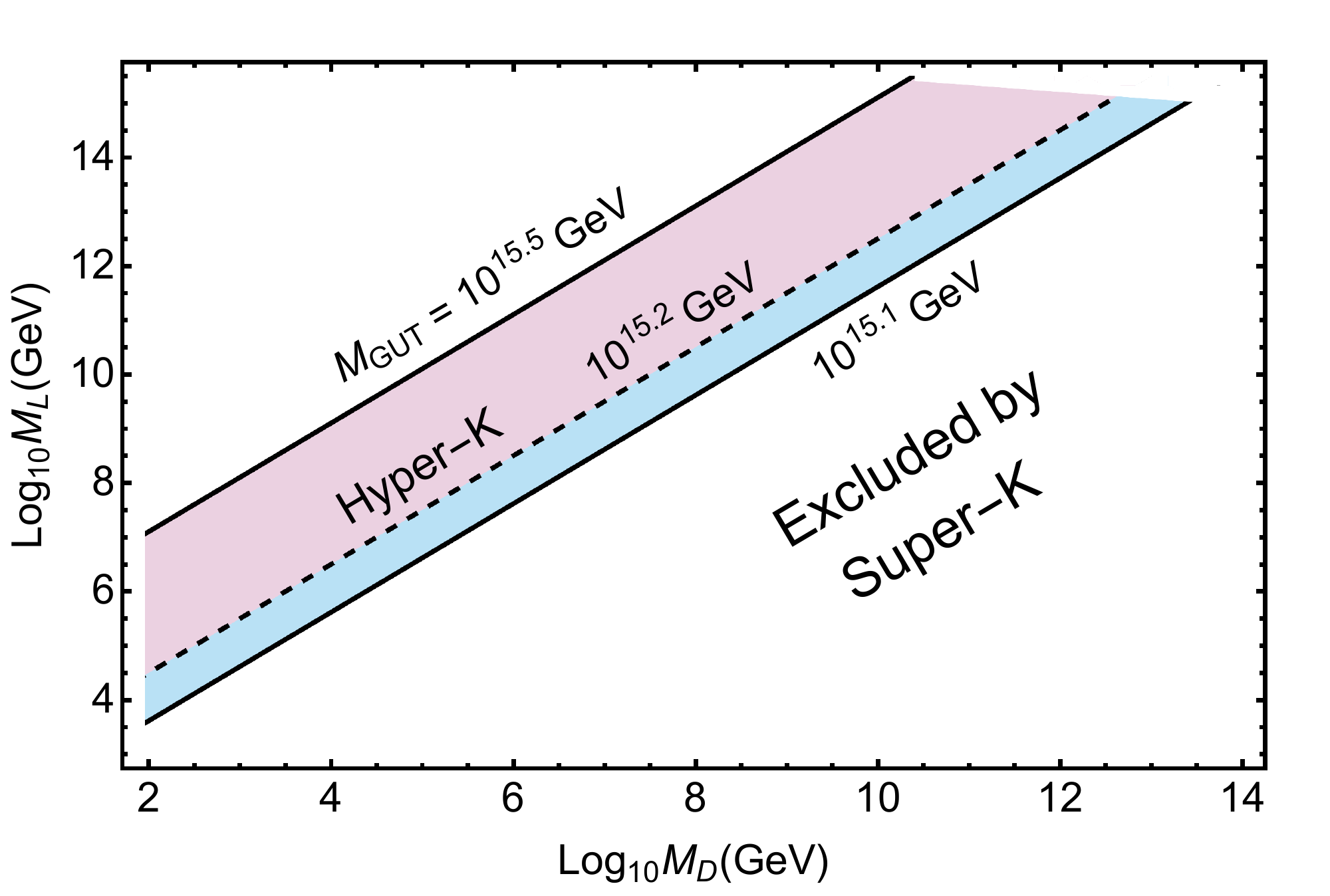}
\caption{Allowed masses of the fields $\rho_3$ and $\rho_8$ consistent with the unification constraints for the most optimistic case: $M_{\Sigma_3}=M_{\Sigma_8}=M_Z$. The white region below the diagonal line is excluded by the bounds on the proton decay lifetime from the Super-Kamiokande collaboration \cite{Abe:2014mwa}.}
\end{figure}

\FloatBarrier

For completeness, in Fig.~3, we show the running of the couplings for a given point in the parameter space allowed by unification consistent with proton decay bounds. As an illustrative example, we choose the scenario corresponding to the maximal GUT scale allowed by the theory.  In the left-panel of Fig.~3 we show the results at one-loop level while in the right-panel we show the results for the maximal GUT scale at two-loop level. Notice that the GUT scale increases approximately in a factor 1.6 when we go from one-loop to two-loop level. This means that the proton lifetime increases in a factor 6.3 approximately. 

\begin{figure}[h]
\includegraphics[width=0.45\linewidth]{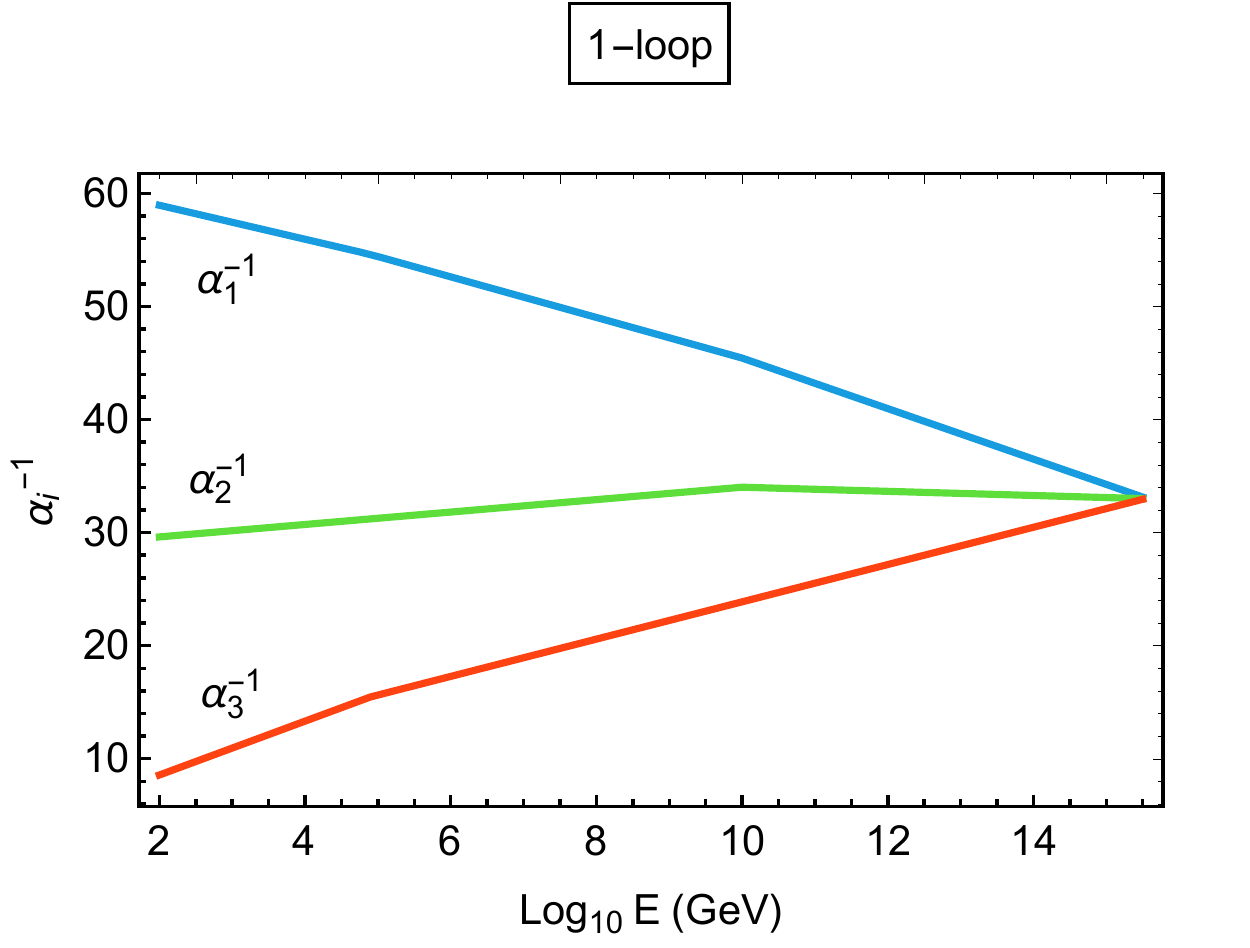}
\includegraphics[width=0.45\linewidth]{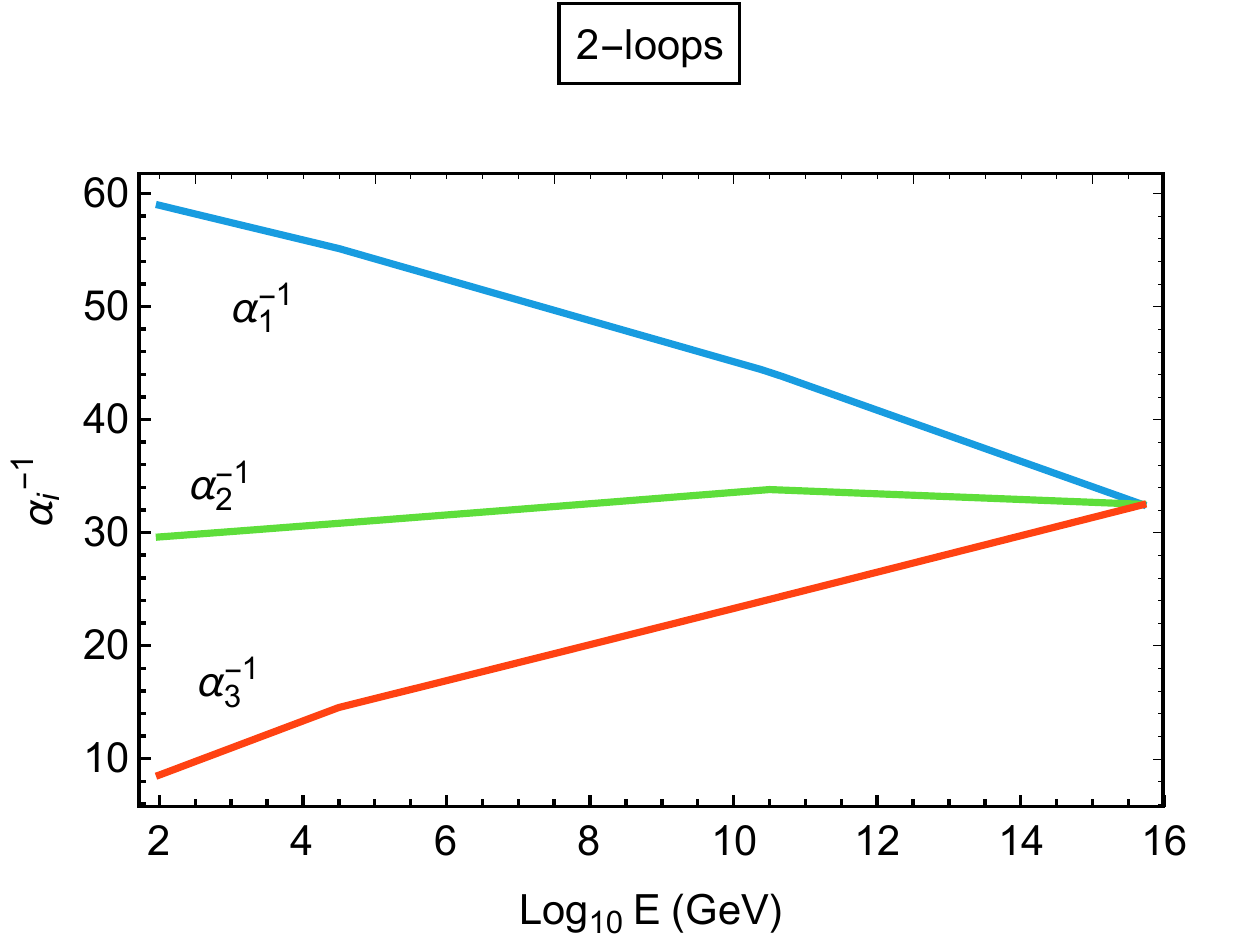}
\caption{On the left panel, running of the couplings at 1-loop level for the scenario where $M_{\Sigma_8}=M_{\Sigma_3}=M_{\rho_3}=M_Z$, $M_{\rho_8}=M_{GUT}$, $M_L=10^{10}$ GeV, $M_D =10^{4.9}$ GeV, and $M_{GUT}=10^{15.5}$ GeV. On the right panel, running of the couplings at 2-loop level for the scenario where $M_{\Sigma_8}=M_{\Sigma_3}=M_{\rho_3}=M_Z$, $M_{\rho_8}=M_{GUT}$, $M_L=10^{10.5}$ GeV, $M_D =10^{4.5}$ GeV, and $M_{GUT}=10^{15.7}$ GeV. }
\end{figure}

\subsection{Proton Decay}
The most dramatic prediction of grand unified theories is the decay of the proton. 
In this theory, it is important to understand if we can satisfy the current proton decay bounds and if we can hope to test its predictions at current or future proton decay experiments.
The relevant decay widths for the proton decay into charged leptons or neutrinos are given by:
\begin{eqnarray}
\displaystyle \Gamma (p \to \pi^0 e^+_\beta)&=& \frac{m_p}{8 \pi} A^2 k_1^4 \left(  \left| c(e^c,d) \matrixel{\pi^0}{(ud)_L u_L}{p}\right |^2  + \left| c(e,d^c) \matrixel{\pi^0}
{(ud)_R u_L}{p}\right|^2 \right),\nonumber \\
\displaystyle \Gamma (p \to K^+ \bar{\nu}) &=& \frac{m_p}{8 \pi} \left( 1- \frac{m_{K^+}^2}{m_p^2}\right)^2 A^2 k_1^4 \sum_i  \left|  
c(\nu_i,d,s^c)  \matrixel{K^+}{(us)_R d_L}{p} + c(\nu_i,s,d^c)  \matrixel{K^+}{(ud)_R s_L}{p} \right|^2,\nonumber\\
\displaystyle \Gamma (p \to \pi^+ \bar{\nu})&=& \frac{m_p}{8\pi} A^2 k_1^4 \sum_i \left | c(\nu_i,d,d^c) \matrixel{\pi^+}{(du)_Rd_L}{p}\right |^2,
\end{eqnarray}
where
\begin{equation}
A=A_{QCD} A_{SR}=\left( \frac{\alpha_3 (m_b)}{\alpha_3 (M_Z)} \right)^{6/23} \left( \frac{\alpha_3 (Q)}{\alpha_3 (m_b)} \right)^{6/25} \left( \frac{\alpha_3 (M_Z)}{\alpha_3 (M_{GUT})} \right)^{2/7},
\end{equation}
where the parameter $k_1=g_{GUT}/\sqrt{2} M_{GUT}$, and $A$ encodes the information for the running of the operators. The numerical values we use are $A_{SR} \approx 1.5$ and $A_{QCD}\approx 1.2$ \cite{Nath:2006ut}.
The matrix elements present in the different decay channels can be computed using lattice QCD. We use the values reported in the recent lattice study~\cite{Aoki:2017puj}:
\begin{align*}
&\matrixel{\pi^0}{(ud)_Lu_L}{p}=0.134(5)(16) \text{ GeV}^2, &\matrixel{\pi^0}{(ud)_Ru_L}{p}=-0.131(4)(13) \text{ GeV}^2, \\
&\matrixel{K^+}{(us)_R d_L}{p}=-0.049(2)(5) \text{ GeV}^2, & \matrixel{K^+}{(ud)_R s_L}{p} =-0.134(4)(14) \text{ GeV}^2, \\
&\matrixel{\pi^+}{(du)_R d_L}{p}=-0.186(6)(18)  \text{ GeV}^2.
\end{align*}
The c-coefficients \cite{FileviezPerez:2004hn} in the above decay channels are given by:
\begin{eqnarray}
&& c(e^c_\alpha, d_\beta) = V_1^{11} V_2^{\alpha \beta} + (V_1 V_{UD})^{1 \beta} (V_2 V_{UD}^\dagger)^{\alpha 1},\\
&& c(e_\alpha, d^c_\beta) = V_1^{11} V_3^{\beta \alpha},\\
&& c(\nu_l,d_{\alpha},d_{\beta}^c) = (V_1V_{UD})^{1\alpha}(V_3 V_{EN})^{\beta l},
\end{eqnarray}
where $\alpha,\beta=1,2$ and $i=1,2,3$. The mixing matrices are defined as:
\begin{equation}
V_1=U_C^{\dagger}U, V_2=E_C^{\dagger}D, \  V_3=D_C^{\dagger}E, V_{UD}=U^{\dagger}D \  \textrm{and}, \  V_{EN}=E^{\dagger}N,
\end{equation}
where the matrices $U, E, D$ and $N$ define the diagonalization of the Yukawa couplings:
\begin{equation}
U_C^T Y_u U=Y_u^{\text{diag}}, \ D_C^T Y_d D=Y_d^{\text{diag}}, \ E_C^T Y_e E=Y_e^{\text{diag}} \ \text{and} \ N^T Y_\nu N=Y_\nu^{\text{diag}}.
\end{equation}
In our theory, $Y_u = Y_u^T$, and this allows us to make a clean prediction for the decay channels into antineutrinos. 
Therefore, the different proton decay channels can be written in a simple way:
\begin{eqnarray}
\Gamma (p \to \pi^0 e^+_\beta)&=& \frac{m_p}{8 \pi} A^2 k_1^4 \left(  \left| (V_2^{11} +V_{UD}^{11}(V_2 V_{UD})^{11}) \matrixel{\pi^0}{(ud)_L u_L}{p} \right |^2 + \left |V_3^{11}  \matrixel{\pi^0}
{(ud)_R u_L}{p}\right|^2 \right),  \nonumber \\
 \Gamma (p \to K^+ \bar{\nu}) &=&   \frac{m_p}{8 \pi} \left( 1- \frac{m_{K^+}^2}{m_p^2}\right)^2 A^2 k_1^4 \left( \left |V_{UD}^{11}  
 \matrixel{K^+}{(us)_R d_L}{p} \right |^2 + \left | V_{UD}^{12}  
 \matrixel{K^+}{(ud)_R s_L}{p} \right|^2 \right), \nonumber \\ 
 \Gamma (p \to \pi^+ \bar{\nu}) &=& \frac{m_p}{8 \pi} A^2 k_1^4\left|V_{UD}^{11} \matrixel{\pi^+}{(du)_R d_L}{p}\right |^2,
\end{eqnarray}
where $V_{UD}^{ij}$ are the elements of the $V_{CKM}$ matrix. We note that one cannot predict the decay width for the channel $p \to e^+ \pi^0$ since we do not 
known the mixing matrices $V_2$ and $V_3$. However, the decay width for the proton decay channels into anti-neutrinos are predicted, and one can use them to define the lower bound on the GUT scale imposing the proton decay experimental bounds.
\begin{figure}[h]
\includegraphics[width=0.55\linewidth]{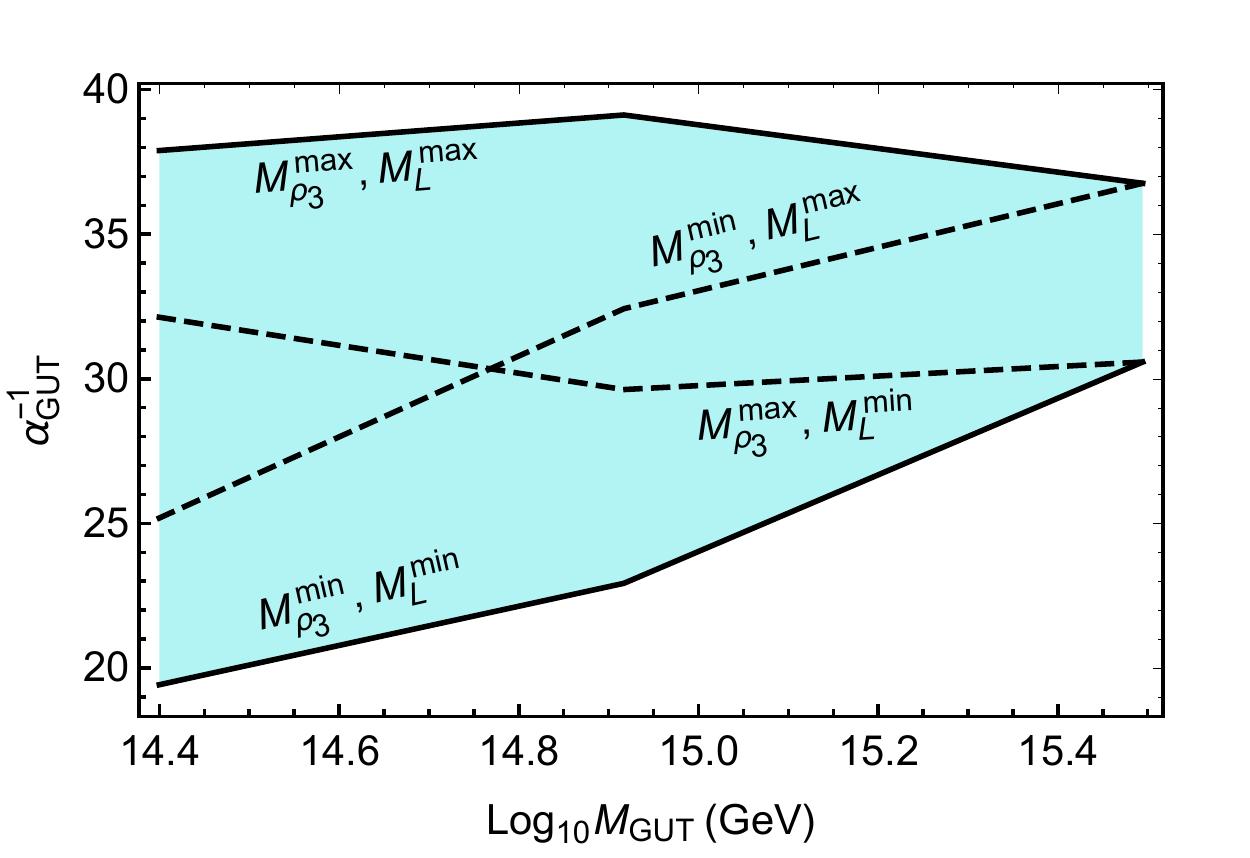}
\caption{Parameter space for the gauge couplings at the GUT scale consistent with unification for the most interesting scenario: $M_{\Sigma_8}=M_{\Sigma_3}=M_{Z}$.}
\end{figure}
\begin{figure}[h]
\includegraphics[width=0.49\linewidth]{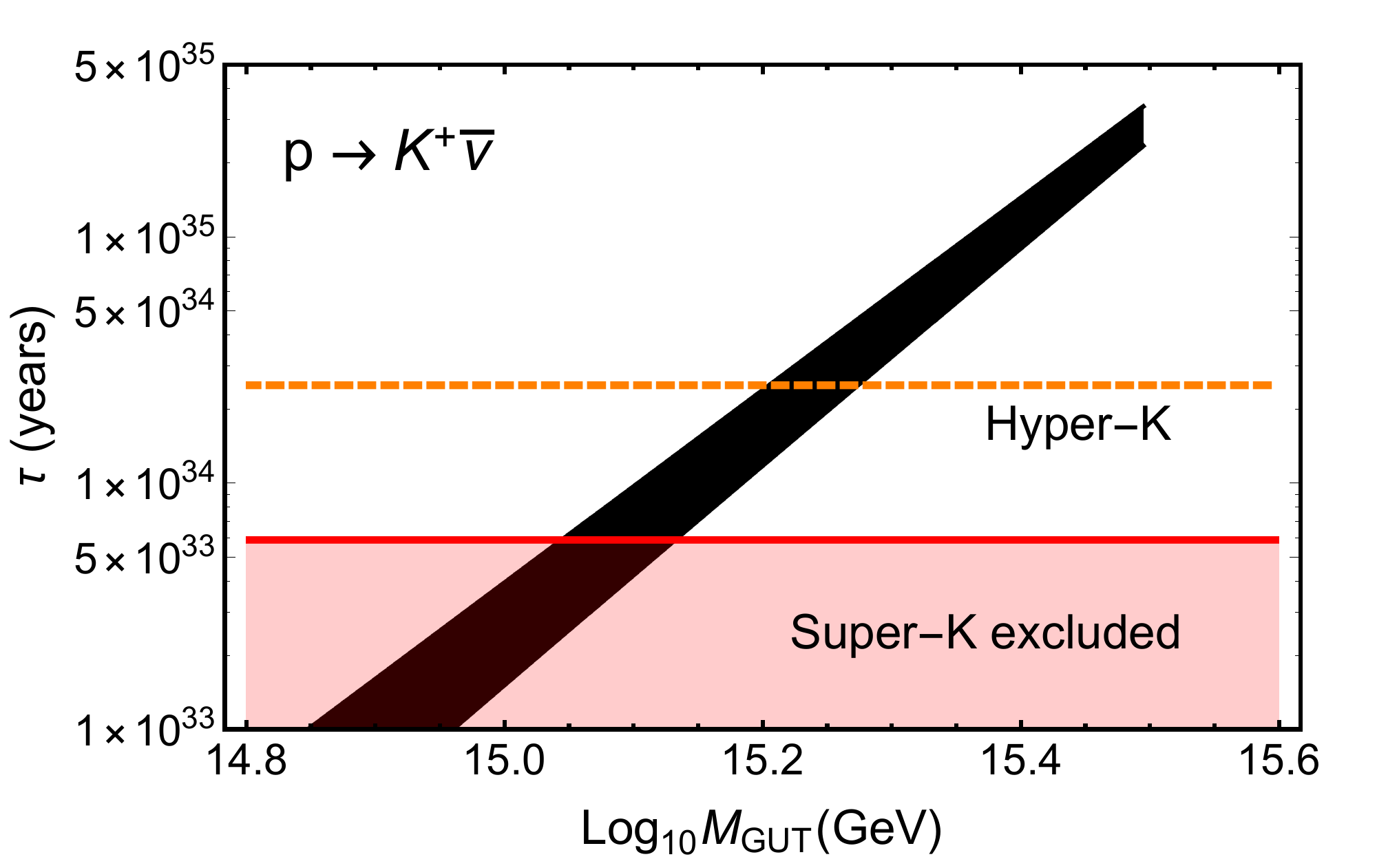}
\includegraphics[width=0.49\linewidth]{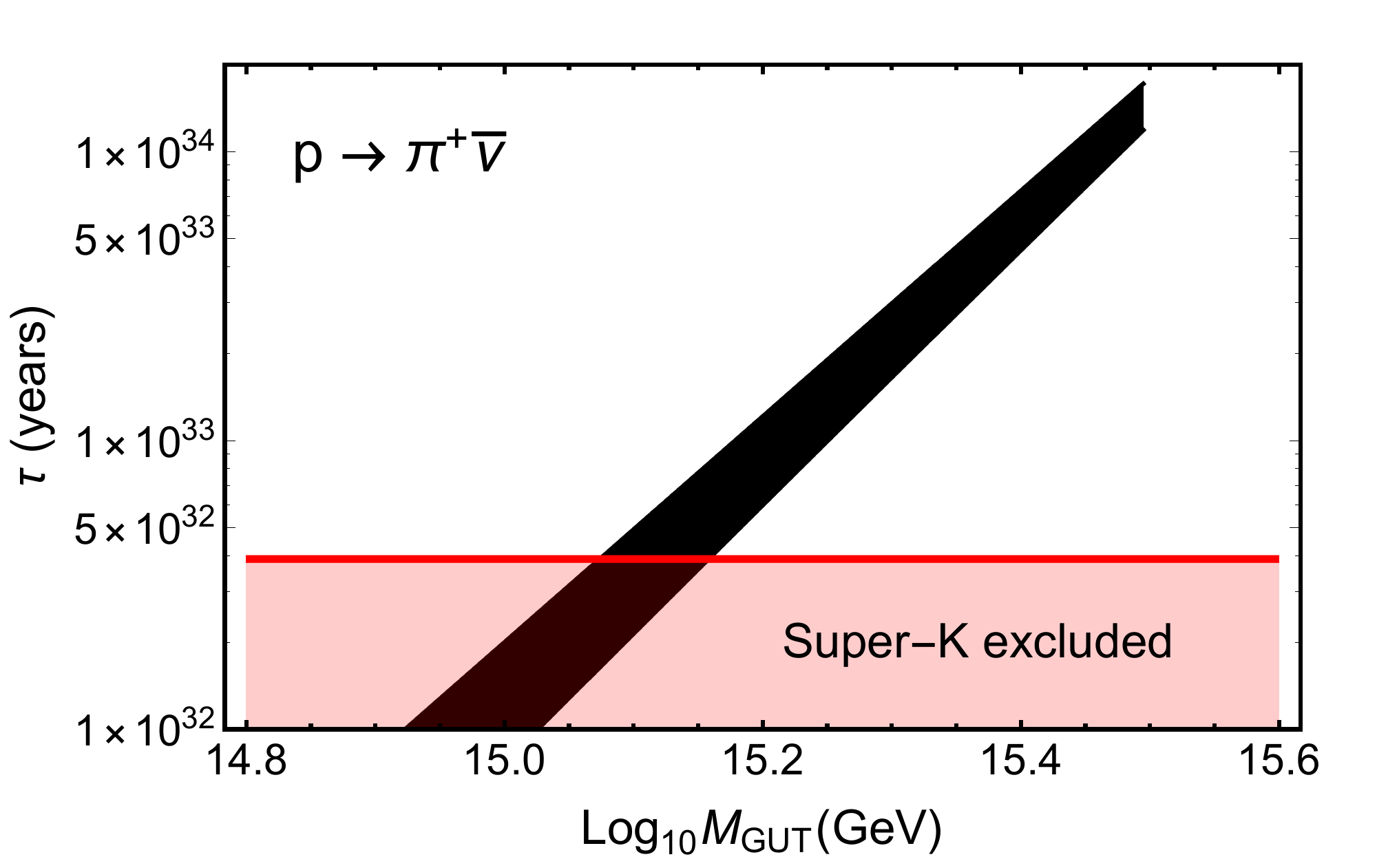}
\caption{Predictions for the proton decay lifetime for the channels $p \to K^+ \bar{\nu}$ (left-panel) and $p \to \pi^+ \bar{\nu}$ (right-panel). The red line corresponds to the current proton decay lifetime for the different channels, 
i.e. $\tau(p \to K^+ \bar{\nu}) > 5.9 \times 10^{33}$ years \cite{Abe:2014mwa} (left-panel) and $\tau(p \to \pi^+ \bar{\nu})> 3.9 \times 10^{32}$ years \cite{Abe:2013lua} (right-panel). The orange dashed line on the left-panel shows the projected bound on $p \to K^+ \bar{\nu}$ from the Hyper-Kamiokande collaboration, i.e. $\tau(p\to K^+ \bar{\nu} > 2.5 \times 10^{34}$ years~\cite{Yokoyama:2017mnt}.}
\end{figure}
\begin{figure}[h]
\includegraphics[width=0.55\linewidth]{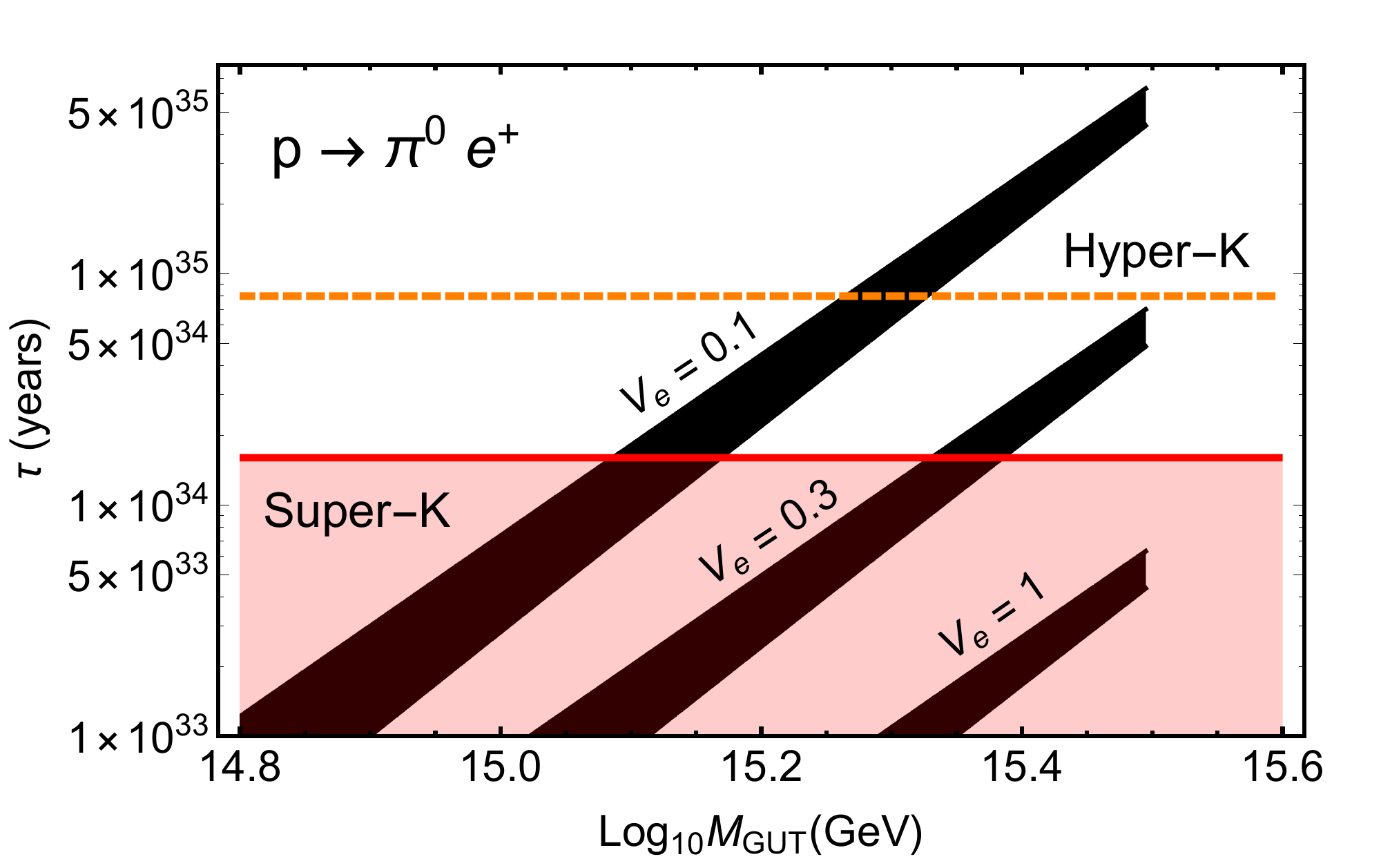}
\caption{Predictions for the proton decay lifetime for the channel $p \to e^+ \pi^0$ using three different values for the unknown mixing $V_e=V_2^{11}=V_3^{11}$. The red line shows the current proton decay lifetime, i.e. $\tau ( p\to \pi^0 e^+) > 1.6 \times 10^{34}$ years \cite{Miura:2016krn}. The orange dashed line shows the projected bound on proton decay lifetime from the Hyper-Kamiokande collaboration, i.e. $\tau(p \to \pi^0 e^+) > 8 \times 10^{34}$ years~\cite{Yokoyama:2017mnt}.}
\end{figure}

In Fig.~4, we show the values for the unified gauge coupling, $\alpha_{GUT}$, in the most interesting scenario allowed by proton decay. Using these results, we can predict the proton decay lifetime for the most relevant proton decay channels. In Fig.~5, we show the predictions for the $p \to K^+ \bar{\nu}$ and $p \to \pi^+ \bar{\nu}$ channels. These results are striking because we can predict an upper bound on proton decay, i.e. $\tau (p \to K^+ \bar{\nu}) \lesssim  3.4 \times 10^{35}$ years and $\tau(p \to \pi^+ \bar{\nu}) \leq 1.7 \times 10^{34}$ years. In Fig.~6, we show the predictions for the $p \to e^+ \pi^0$ channel, but since the theory does not predict the relevant mixing matrices, we cannot make a strong prediction. We emphasize that in contrast to other GUT theories, where the correction of the fermion mass relations sacrifices the prediction $Y_u = Y_u^T$, we find clean channels which do not depend on any unknown mixing matrix. This allows us to set an upper bound on the proton decay lifetime; thus, there is hope to test this theory in future proton decay experiments. 
\subsection{Neutrino Masses}
It is important to show that one can generate at least two massive neutrinos in the context of this theory.
The relevant terms in the Lagrangian for our discussion are given by:
\begin{equation}
 -{\cal L}\supset   \frac{v_5}{2\sqrt{2}}  \left(y_0^i \,  \nu_i   +  y_1^i \, N_i - y_2^i \, N_i^c \right)  \left(\rho_3 + \xi \, \rho_0 \right)  - N M_L N^c + \frac{1}{2} M_{\rho_0}\rho_0^2 + \frac{1}{2}M_{\rho_3} \rho_3^2 
- \nu \tilde{M}_{\bar{5}5} N^c  + \rm{h.c.} ,
\end{equation}
where $\xi =3/\sqrt{15}$, $\tilde{M}_{\bar{5}5}= M_{\bar{5}5} - 3 \lambda_{\bar{5}5} M_{GUT}/(2\sqrt{25\pi \alpha_{GUT}})$, and $v_5$ is the vacuum expectation value of the Standard Model Higgs.

\begin{figure}[h]
\includegraphics[width=0.3\linewidth]{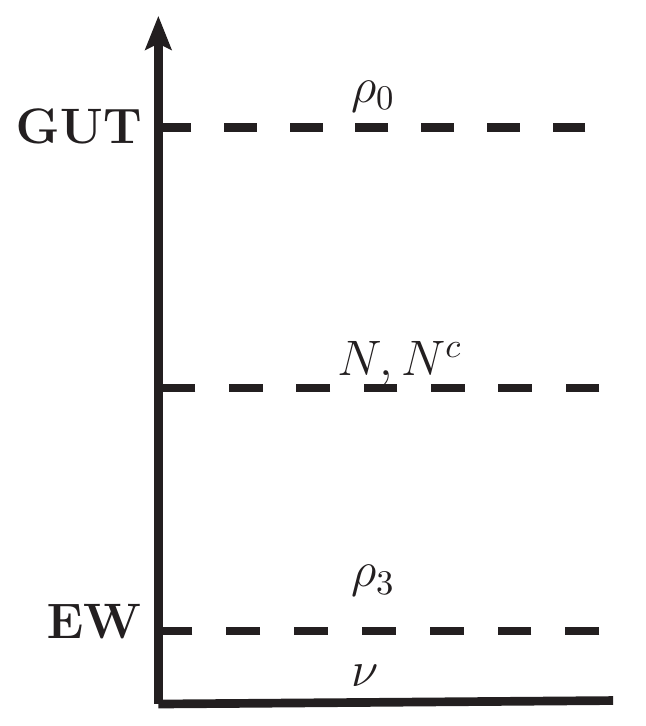}
\caption{Hierarchy of the neutral fermionic fields required by unification and proton decay constraints.}
\end{figure}
The achievement of unification together with proton decay bounds imposes a well-defined hierarchy in the neutral fermionic sector; see Fig.~7. As we have shown, the minimum mass splitting in  {\bf 24} required by proton decay bounds is nine orders of magnitude, which places $\rho_3$ and $\rho_0$ near the electroweak and GUT scale, respectively. As we show in Fig.~2, one has more freedom for the mass splitting in ${\bf 5}$ and ${\bf 5^{'}}$ representations, and the vector-like fermions could live anywhere in the great desert. 

According to the established hierarchy, $M_{\rho_0} \gg M_L \gg M_{\rho_3} \gg M_\nu$, we first integrate out the heaviest neutral field, $\rho_0$. This generates a contribution to the neutrino mass and some mixing terms between $N$ and $N^c$, all suppressed by $M_{\rho_0}$. At this point, the mass matrix for the $N$ and $N^c$ fields is:
\begin{equation}
\displaystyle \begin{pmatrix} N_i && N^c_i \end{pmatrix} \begin{pmatrix} - \frac{1}{4} y_1^iy_1^j  \, \xi^2 \, v_5^2 /M_{\rho_0} && - M_L^{ij}  \\ -M_L^{ji} &&- \frac{1}{4} y_2^i y_2^j\, \xi^2 \,  v_5^2/M_{\rho_0}\end{pmatrix}\begin{pmatrix} N_j \\ N^c_j \end{pmatrix},
\end{equation}
where we keep only the lowest order terms. The $N$ and $N^c$ fields can be written as a linear combination of the new fields:
\begin{eqnarray}
N&=& \cos \theta N_1 + \sin \theta N_2, \\
N^c &=& -\sin \theta N_1 + \cos \theta N_2,
\end{eqnarray}
where the mixing angle $\theta$ is defined by the diagonalization of the mass matrix:
\begin{equation}
\tan 2\theta= \frac{ 8 M_L^{ij} M_{\rho_0}}{(y_2^iy_2^j-y_1^i y_1^j)  \, \xi^2 v_5^2}.
\end{equation}
Clearly, since $M_L M_{\rho_0} \gg v_5^2$, the mixing angle is $\theta \sim \pi / 4$, and the eigenvalues are $\pm M_L$. Note that we can always rotate the field $N_2 \to N_2 e^{i\frac{\pi}{2}}$ to define positive masses. According to the mass hierarchy in the neutral fermions, we can integrate $N_1$ and $N_2$ out, which leads to the following effective Lagrangian for the light degrees of freedom: 
\begin{equation}
-{\cal L}^{ \nu,\rho_3}_{eff} \supset  -\left( \xi^2 \, y_0^i y_0^j \frac{v_5^2}{8M_{\rho_0}} \right) \nu_i \nu_j + \frac{1}{2}M_{\rho_3} \rho_3^2+ \frac{v_5}{2\sqrt{2}}\nu_i \big( y_0^i-(y_1^TM_L^{-1}\tilde{M}_{\bar{5}5}^T)^i \big)\nu_i\rho_3,
\end{equation}
where we do not include terms of order $(M_LM_{\rho_0})^{-1}$ and we neglect corrections to the $\rho_3$ mass suppressed by $M_L$. We note that in the limit $\theta \to \pi/4$, there is no contribution to the light neutrino mass term from the vector-like leptons; however, we point out that they do contribute to the effective coupling between $\nu$ and $\rho_3$. This will be a key point to predict a consistent neutrino mass spectrum.

Finally, by integrating out $\rho_3$, the final seesaw takes place, and a one generates a new contribution to the light neutrinos suppressed by $M_{\rho_3}$. The effective mass term for the light neutrinos is given by:
\begin{equation}
M_{\nu}^{ij}=  \frac{v_5^2 }{4} \left( \frac{\xi^2 \, y_0^iy_0^j}{M_{\rho_0}}  + \frac{\big(y_0^i -  (y_1^TM_L^{-1}\tilde{M}_{\bar{5}5}^T)^i\big) \big(y_0^j-(y_1^TM_L^{-1}\tilde{M}_{\bar{5}5}^T)^j \big)}{M_{\rho_3}} \right).
 \end{equation}
We summarize below the different contributions to the light neutrino mass term in a schematic way:
\begin{equation}
\includegraphics[width=0.8\linewidth]{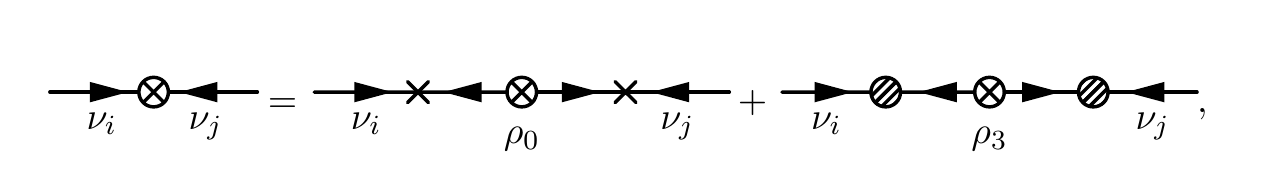}
\end{equation}
where $\times$ symbolises a tree-level interaction between the fermions, ${\bf \otimes}$ represents a mass term insertion, and we define the effective vertex:
\begin{equation}
\includegraphics[width=0.8\linewidth]{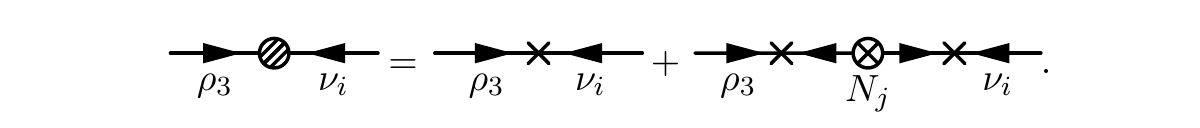}
\end{equation}

We note that with only the contribution of the ${\bf 24}$, the model would predict two massless neutrinos; thus, the presence of the ${\bf 5^{'}}$ and ${\bf \bar{5}^{'}}$ is crucial to guarantee the consistency of the theory with experiment. 
\subsection{Charged Fermion Masses}
As we discussed above, one of the main problems of the Georgi-Glashow model is that 
one predicts $Y_e=Y_d^T$. In this theory, we can achieve a consistent relation 
between the masses for charged leptons and down quarks. We can compute the masses 
using the following terms:
\begin{eqnarray}
- {\cal L} &\supset & \, \,E M_LE^c +  D^c M_D D +M_{\rho_3}\rho_3^+\rho_3^-+d^c \left( M_{\bar{5}5} +  \tilde{\lambda}_{\bar{5}5}\frac{M_{GUT}}{\sqrt{\alpha_{GUT}}} \right)  D +
 e \left( M_{\bar{5}5}   -\frac{3}{2} \tilde{\lambda}_{\bar{5}5}\frac{M_{GUT}}{\sqrt{\alpha_{GUT}}}  \right) E^c \nonumber \\
 && + \frac{v_5}{\sqrt{2}} \left( e Y_1 e^c + d^cY_1 d +D^cY_2 d +E Y_2e^c \right) +   \frac{v_5}{2} \left( y_0^i e_i \rho_ 3^+ + y_1^i \,  E_i \rho_ 3^+ + y_2^i \,  \rho_3^- E^c_i\right) ,
\end{eqnarray}
where $\tilde{\lambda}_{\bar{5}5}=\lambda_{\bar{5}5}/\sqrt{15\pi}$. We find the following mass matrix for the down-type quarks:
\begin{equation}
\begin{pmatrix} d^c &&  D^c \end{pmatrix} \begin{pmatrix} \displaystyle Y_1 \frac{v_5}{\sqrt{2}} && \displaystyle M_{\bar{5}5} +\tilde{\lambda}_{\bar{5}5}\frac{M_{GUT}}{\sqrt{\alpha_{GUT}}} \\ \displaystyle Y_2 \frac{v_5}{\sqrt{2}} && M_D \end{pmatrix} \begin{pmatrix} d \\ D \end{pmatrix},
\end{equation}
where we have neglected the mixing proportional to $d^c \rho_{(3,2)}$ and $D^c \rho_{(3,2)}$ since they enter in the light neutrino mass matrix. The mass matrix for the charged leptons is given by:
\begin{equation}
\begin{pmatrix} e^c && E^c && \rho_3^+ \end{pmatrix} \begin{pmatrix} \displaystyle Y_1^T \frac{v_5}{\sqrt{2}}  && \displaystyle Y_2^T \frac{v_5}{\sqrt{2}} && 0 \\ \displaystyle M_{\bar{5}5}^T-\frac{3}{2} \tilde{\lambda}^T_{\bar{5}5}\frac{M_{GUT}}{\sqrt{\alpha_{GUT}}} && M_L^T &&\displaystyle  \frac{1}{2}y_2 \, v_5 \\ \displaystyle  \frac{1}{2}y_0 \, v_5 && \displaystyle \frac{1}{2} y_1 \, v_5 && M_{\rho_3} \end{pmatrix} \begin{pmatrix} e \\ E \\ \rho_3^- \end{pmatrix}.
\end{equation}
Clearly, there is enough freedom to have a consistent relation between the masses of the charged leptons and down quarks. We refer the reader to Ref.~\cite{Dorsner:2014wva} for a detailed study on the role of ${\bf 5^{'}}$ and ${\bf \bar{5}^{'}}$ representations in the achievement of realistic charged fermion masses at the low scale.

\FloatBarrier

\section{Summary}
We have investigated a simple, realistic grand unified theory based on $SU(5)$ 
where one can generate fermion masses consistent with experiment and
predict an upper bound on proton decay for the channels with antineutrinos:
$\tau(p \to K^+ \bar{\nu}) \lesssim 3.4 \times 10^{35}$ years and $\tau (p \to \pi^+ \bar{\nu}) \lesssim 1.7 \times 10^{34}$ years. 
In this context, we can have 
a consistent relation between the charged lepton and down quark masses 
due to the presence of the new vector-like fermions. The neutrino masses are 
generated through the type I and type III seesaw mechanisms, and we find that 
the field responsible for the type III seesaw mechanism must be light, i.e. 
$M_{\rho_3} \lesssim 500$ TeV. This theory can be considered as one of the 
appealing candidates for unification based on $SU(5)$, as it can be tested 
in current or future proton decay experiments.     \\

{\textit{Acknowledgments}}: The work of P.F.P. has been supported by the U.S. Department of Energy under contract No. de-sc0018005. The work of C.M. has been supported in part by the Spanish Government and ERDF funds from the EU Commission [Grants No. FPA2014-53631-C2-1-P and SEV-2014- 0398] and "La Caixa-Severo Ochoa" scholarship. C.M. thanks Case Western Reserve University for the great hospitality. P. F. P. thanks the Walter Burke Institute for Theoretical Physics at Caltech for hospitality.
\appendix

\section{RGE of the gauge couplings at two-loops}
The RGEs at two-loop level can be written as
\begin{equation}
\frac{d \, \alpha_i (\mu)}{d\, \text{ln} \mu}=\frac{b_i}{2\pi}\alpha_i^2(\mu) + \frac{1}{8\pi^2}\sum_{j=1}^3b_{ij} \, \alpha_i^2(\mu) \, \alpha_j(\mu)+\frac{1}{32\pi^3} \, \alpha_i^2(\mu)\sum_{\ell = U,D,E}\text{Tr} [C_{i\ell}Y_\ell^\dagger Y_\ell ]
\end{equation}
where $\alpha_i = g_i^2/4 \pi$ and the $Y_\ell$ are the Yukawa couplings. The $b_i$ and $b_{ij}$ are given by
\begin{eqnarray*}
&&b_i^{d^c}=b_i^{D^c}=\begin{pmatrix} \frac{2}{15} \\ 0 \\ \frac{1}{3} \end{pmatrix}, \,\,\, b_i^{e^c}=\begin{pmatrix} \frac{2}{5} \\ 0 \\ 0  \end{pmatrix}, \,\,\, b_i^{u^c}=\begin{pmatrix} \frac{8}{15} \\ 0 \\ \frac{1}{3} \end{pmatrix}, \,\,\,b_i^{q}=\begin{pmatrix} \frac{1}{15} \\ 1 \\ \frac{2}{3} \end{pmatrix}, \,\,\, b_i^{H}=\begin{pmatrix} \frac{1}{10} \\ \frac{1}{6} \\ 0 \end{pmatrix}=\frac{1}{2} b_i^{\ell}, = \frac{1}{2} b_i^L \,\,\,  \\
&&b_i^{\Sigma_3}=\begin{pmatrix} 0 \\  \frac{1}{3} \\ 0 \end{pmatrix}, \,\,\, b_i^{\Sigma_8}=\begin{pmatrix}  0 \\ 0 \\ \frac{1}{2} \end{pmatrix},\,\,\, b_i^{\rho_3}=\begin{pmatrix} 0 \\  \frac{4}{3} \\ 0 \end{pmatrix}, \,\,\, b_i^{\rho_{32}}=\begin{pmatrix}  \frac{5}{3} \\ 1 \\ \frac{2}{3} \end{pmatrix}, \,\,\,b_i^{\rho_8}=\begin{pmatrix} 0 \\ 0 \\ \frac{4}{3} \end{pmatrix}
\end{eqnarray*}
Notice that here we show the contributions of only one family of the SM fields.
\begin{eqnarray*}
&&b_{ij}^{d^c(D^c)}=\begin{pmatrix} \frac{2}{75} && 0 && \frac{8}{15} \\ 0 && 0 && 0 \\ \frac{1}{15} && 0 && \frac{19}{3} \end{pmatrix}, \,\,\, b_{ij}^{H}=\begin{pmatrix} \frac{9}{50} && \frac{9}{10} && 0 \\ \frac{3}{10} && \frac{13}{6} && 0 \\ 0 && 0 && 0 \end{pmatrix},\,\,\, b_{ij}^{q}=\begin{pmatrix} \frac{1}{300} && \frac{3}{20} && \frac{4}{15} \\ \frac{1}{20} && \frac{49}{4} && 4 \\ \frac{1}{30} && \frac{3}{2} && \frac{38}{3} \end{pmatrix}, \,\,\, b_{ij}^{u^c}=\begin{pmatrix} \frac{32}{75} && 0 && \frac{32}{15} \\ 0 && 0 && 0 \\ \frac{4}{15} && 0 && \frac{19}{3}  \end{pmatrix}, \\
&& b_{ij}^{e^c}=\begin{pmatrix} \frac{18}{25} && 0 && 0 \\ 0 && 0 && 0 \\ 0 && 0 && 0 \end{pmatrix}, \,\,\, b_{ij}^{\ell(L)}=\begin{pmatrix} \frac{9}{100} && \frac{9}{20} && 0 \\ \frac{3}{20} && \frac{49}{12} && 0 \\ 0 && 0 && 0 \end{pmatrix},  \,\,\, b_{ij}^{\rho_{32}}=\begin{pmatrix} \frac{25}{12} && \frac{15}{4} && \frac{20}{3} \\ \frac{5}{4} && \frac{49}{4} && 4 \\ \frac{5}{6} && \frac{3}{2} && \frac{38}{3} \end{pmatrix},  \,\,\, b_{ij}^{\rho_8}= \begin{pmatrix} 0 && 0 && 0 \\ 0 && 0 && 0 \\ 0 && 0 && 48 \end{pmatrix}, \\
&&\,\,\,  b_{ij}^{\rho_3}= \begin{pmatrix} 0 && 0 && 0 \\ 0 &&  \frac{64}{3} && 0 \\ 0 && 0 && 0 \end{pmatrix}, \,\,\, b_{ij}^{\Sigma_8}= \begin{pmatrix} 0 && 0 && 0  \\ 0 && 0 && 0 \\ 0 && 0 && 21 \end{pmatrix}, \,\,\, b_{ij}^{\Sigma_3}=\begin{pmatrix} 0 && 0 && 0 \\ 0 && \frac{28}{3} && 0 \\ 0 && 0 && 0 \end{pmatrix}.
\end{eqnarray*}
Here we follow and use the notation of the Ref.~\cite{REGs}.
%


\end{document}